\documentclass[12pt,a4paper]{article}
\usepackage{amsfonts,amsmath}
\usepackage{epsfig}

\def \ep {\epsilon}
\def \g {\varepsilon}

\def\oI{\mathbb{I}}
\def\ICP{\mathbb{CP}}

\def \tr {{\rm tr}}
\def \ha {{1 \over 2}}
\def \td {\tilde}
\def \ci{\cite}
\def \N {{\mathcal N}}

\def \k {\kappa}
\def\foot{\footnote}

\def \ci {\cite}
\def \ov {\over}
\def \bp {\begin{pmatrix}}  \def \epm {\end{pmatrix}}
\def \ha {{\textstyle{1 \ov 2}}}
\def \bi{\bibitem}
\def \la {\label}
\def \l {\lambda}
\def\foot{\footnote}

\newcommand{\rf}[1]{(\ref{#1})}
\def \ci {\cite}
  
\def \m {\mu}
\def \m   {\mu}
\def \by {\times}

\def \n {\nu}
\def \bY {\bar Y}
  
  \def \JJ  {{\rm J}}
    \def \AA  {{\rm A}}

\def \no {\nonumber}
\def \del {\partial}
\def \be {\begin{eqnarray}}
\def \ee {\end{eqnarray}}

\begin{document}

% \hoffset -0.35 cm \textheight 21.0 cm \textwidth 15.2 cm \topmargin
% -0.54 cm \oddsidemargin 0.8 cm
% \tolerance=300% eto bor'ba s overfullom
% \hfuzz=2.pt  % i eto bor'ba s overfullom

\renewcommand{\theequation}{\thesection.\arabic{equation}}
\renewcommand{\thesubsection}{\arabic{section}.\arabic{subsection}}
\makeatletter \@addtoreset{equation}{section} \makeatother

\newcommand{\bZ}{{\bar Z}}
\newcommand{\Tr}{{\rm Tr}}
\newcommand{\A}{{\cal A}}
\newcommand{\C}{\mathbb C}
\newcommand{\Z}{\mathbb Z}

 \def \g {\varepsilon}

  \def \lb {L-BLG\ }
  \def \a {\alpha}

\def \s {\sigma}

\thispagestyle{empty}
\begin{flushright}\footnotesize
%\texttt{AEI-2008-077}
\texttt{ Imperial-TP-EA-2008-1 }
\vspace{0.8cm}
\end{flushright}

\renewcommand{\thefootnote}{\fnsymbol{footnote}}
\setcounter{footnote}{0}

\begin{center}
{\Large\textbf{\mathversion{bold}
On 3d ${\cal N}$=8 Lorentzian
BLG theory \\
\vspace{0.1cm}
 as a scaling  limit of \\
 \vspace{0.1cm}
 3d superconformal ${\cal N}$=6  ABJM theory
}\par}

\vspace{1.5cm}

{E. Antonyan\footnote{Also at the
Institute for Theoretical Physics and Modeling, Yerevan, Armenia}
  and A.A. Tseytlin\footnote{Also
at the Lebedev Institute,  Moscow.} } \vspace{8mm}

 \textit{ Theoretical Physics Group \\
   The Blackett Laboratory, Imperial College,\\
London SW7 2AZ, U.K.       }
 \vspace{3mm}
 \end{center}

\par\vspace{1cm}

 \begin{abstract}
 We elaborate on the suggestion made in arXiv:0806.3498
 that the 3d ${\cal N}=8$ superconformal  $SU(N)$ Chern-Simons-matter
  theory of  ``Lorentzian''  Bagger-Lambert-Gustavson type (L-BLG)
 can be obtained by a scaling limit (involving sending the level $k$ to
 infinity and redefining the fields) from the
 ${\cal N}=6$ superconformal  $U(N)\times U(N) $ Chern-Simons-matter
 theory of Aharony, Bergman, Jafferis  and Maldacena (ABJM).
 We show that to  implement such  limit in a consistent way one is to extend
 the ABJM theory by an    abelian ``ghost'' multiplet.
 The corresponding limit at the 3-algebra level   also requires
 extending the non-antisymmetric Bagger-Lambert 3-algebra
 underlying  the ABJM theory by a negative-norm generator.
 We draw analogy with similar scaling limits discussed
 previously for  bosonic  Chern-Simons theory
 and  comment on  some   implications of
  this   relation between the ABJM and L-BLG theories.
 \end{abstract}
\newpage

%\vspace{0.5cm}

%%%%
%\setcounter{page}{1}
\renewcommand{\thefootnote}{\arabic{footnote}}
\setcounter{footnote}{0}

%\tableofcontents

%%%%%%%%%%%%%%%%%%%%%%%%%%%%%%%%%%%%%%%%%%%%%%%%%%%%%%%%%%%%%%%%%%%%%%%%%%%

\section{Introduction}

An important problem that attracted  much recent attention
is  to understand a low-energy limit
of a hypothetical  theory  that generalizes the
 worldvolume  theory of a single M2-brane \ci{bst}
to the case of $N$ coincident M2-branes.
     Using some earlier ideas of \ci{harv},
    Bagger and Lambert, and Gustavson (BLG)
  succeeded in constructing  a three-dimensional ${\cal N}=8$ superconformal Chern-Simons-matter
    theory  based on a 3-algebra  \ci{bg1,gu,bg2}.
     However,   it was soon realized  that the original BLG theory can
    describe only two  coincident M2-branes if the 3-algebra is kept
    finite and having  a positive-definite metric \ci{ramsd,papa,med}.

    An interesting suggestion of how to construct a  similar
     theory with right
    symmetries  and degrees of freedom
    to describe $N$  M2-branes was  made in
    \cite{gom,ben,hom} (see also \ci{sch,gomi,mukh2,ver})
    where  a 3-algebra with  a  Lorentzian (indefinite)  signature  metric
    was used.
    The resulting Lorentzian-BLG (L-BLG)  theory
    is  formally ${\cal N}=8$   superconformal at the classical level, but its
    interpretation as a quantum theory appears to be non-trivial
    (in particular, as there is no obvious expansion parameter and its perturbative definition
     depends  on a choice of a vacuum). If expanded near an obvious classical
    vacuum that spontaneously breaks the    superconformal symmetry, it becomes equivalent
    \ci{ben,sch,mukh2,mukh1}
    to the standard low-energy gauge theory of  multiple D2-branes (non-conformal  ${\cal N}=8$
     supersymmetric 3d SYM theory), i.e.
     it  may thus be viewed  as a  conformal ``dressing''
     of non-conformal  3d  ${\cal N}=8$  SYM theory.\foot{There  is a close  analogy
     with  a conformal plane-wave  2d  sigma model \ci{tse} having   $D+2$ dimensional  Lorentzian
     target space $ds^2_{D+2} = dx^+ dx^- +   G_{ij} (x^+, x) dx^i dx^j$
        which  may be viewed as a ``dressing'' of  non-conformal
     2d sigma model with the Euclidean $D$ dimensional target space   metric
      $ds^2_{D}  =  G_{ij} (x) dx^i dx^j$   that depends on $x^+$   according to the RG equation
     ${ \del  G_{ij}\ov  \del x^+ }
     = \beta_{ij} (G)$.
     }

    A different 3d superconformal Chern-Simons-matter   theory was proposed
    by Aharony, Bergman, Jafferis and Maldacena (ABJM)
     \cite{aha}; it  has an  explicit ${\cal N}=6$ supersymmetry  \ci{benak}
      and  may be interpreted as   describing  $N$ coincident
      M2-branes at the singularity of the orbifold $\C^4/\Z_k$.
      For $N=2$ it should be equivalent to the BLG  theory  in the $SU(2) \times SU(2)$
      formulation \ci{ramsd}.

While the ABJM theory also admits  a 3-algebra  interpretation \ci{bag}
%AT
(where  antisymmetry condition is relaxed, see also \ci{st})
it appears to be very  different from the  L-BLG one:
the two theories have different  field content and different symmetries.

In  ref.\ci{hon} an interesting suggestion was  made
that  the L-BLG theory  may  be interpreted as a certain limit  of the ABJM theory,
in which  one sends  the ABJM coupling $k$ (CS level) to infinity
and at the same time  rescales some of the  fields to zero so that they decouple.
However, it was  not  shown   in \ci{hon}  that such  limit of the ${\cal N}=6$
 ABJM action does indeed  lead to the full ${\cal N}=8$ supersymmetric  L-BLG action.

\

%%%%%%%%%%%%%%%%%%%%%%%%%
Here we shall refine the suggestion of \ci{hon}  by
pointing out that  to be able to relate the two theories by a scaling limit
one needs to supplement the ABJM theory with an extra abelian ``ghost'' multiplet
(decoupled from the  ABJM fields). We shall
demonstrate that there exists  a limit (or ``contraction'') of the 3-algebra  underlying
the ABJM theory  \ci{bag} trivially extended by an extra ``ghost'' generator
that leads exactly to the Lorentzian 3-algebra
underlying the L-BLG theory \cite{gom,ben,hom}.
The BLG construction of the
superconformal theory from a 3-algebra  then guarantees that  the corresponding limit of the
(extended) ABJM action is indeed the full  L-BLG action.
This relation between 3-algebras  reinforces the idea of \cite{bag} that 3-algebras
   may be  an essential part of the multiple M2-brane theory.

The scaling limit we  shall consider
 is very  similar to the  limits  considered previously for 2d  WZW models \ci{nap,sfets}, 3d
Chern-Simons  and 4d Yang-Mills theories \ci{tse2}
where one goes (via an infinite ``boost'' in field space and a rescaling of coupling)
from a theory defined  by a  product of a ``ghost'' (time-like) direction  and a simple Lie group
to a theory defined by
 non-semisimple contraction of the corresponding algebra
with  non-degenerate (but  indefinite) metric.\foot{In 2 dimensions  this
limit is a  special case  of a  ``Penrose-type'' limit that leads  to a  plane-wave-type sigma model
    \ci{sfets}. The infinite rescaling of the overall coefficient of the action (or string tension)
  which is part of the  limiting procedure  implies that the resulting model is conformally invariant.
 Let us note also that the  enhancement of supersymmetry in Penrose limit is a well-known phenomenon.}

In the 3d  CS example  considered in \ci{tse2}   the
 starting theory   was the $U(1)_{-k} \times  SU(2)_k $  Chern-Simons one
  and the limiting theory was the CS model
 based on the centrally extended Euclidean algebra $E_2^c$ in 2 dimensions. Here we shall start with
 the  $  [U(N)_k \times U(N)_{-k}]$ Chern-Simons-matter
  ${\cal N}=6$ ABJM theory \ci{aha,bag} with
   an extra  decoupled ``ghost'' multiplet
 and end up with  the $SU(N)$   Chern-Simons-matter ${\cal N}=8$ L-BLG theory
  \ci{gom,ben,hom}.

 One motivation for  considering this limit is that  it may be possible to view it
  as a definition
 of the L-BLG theory in terms of the ABJM theory  and that
  may shed light on the interpretation of the former.
In particular, we shall see how the relation  between the
 L-BLG theory and the  3d $\N=8$ SYM (or D2-brane) theory can be
understood in the  ABJM framework:
 taking the scaling limit and then giving one of the scalars
     an expectation value is actually equivalent to the Higgs-type
     procedure of \cite{mukh1,dis} for obtaining the  D2-brane theory from the ABJM theory.

\

    We shall start in section 2
    with  defining the relevant 3-algebras and reviewing the ABJM and L-BLG theories.
    After  reviewing in section  3.1  the scaling limit in the bosonic CS  theory
    we shall then  explain in sect 3.2   how to relate
    the kinetic terms  in  the ABJM action  combined with a ``ghost'' multiplet
    to those of the L-BLG theory  by a similar scaling limit.
     We will  proceed to show in section 3.3
      how the scaling limit transforms the ABJM
     3-algebra ``tensored'' with a  ghost generator
     into the  Lorentzian BLG 3-algebra.
     We shall draw some  conclusions   in section 4,  commenting  in particular
     on the relation of the two theories to the  D2-brane theory.

\section{Review of ABJM and L-BLG theories}

Let  us start with a  review of basic definition of  3-algebra
that can be used \ci{bag}  to construct the interaction terms in the  supersymmetric
CS-matter  theories like ABJM and L-BLG.

    A 3-algebra is a (complex)
    vector space with a basis $T^a$, $a=1,\ldots,M$, endowed with a triple product
    \begin{equation}
        [T^a,T^b;T^c]=f^{abc}{}_d T^d \ ,
    \end{equation}
    where the structure constants satisfy the following fundamental identity
    \begin{equation}\label{FI}
        f^{efg}{}_bf^{cba}{}_d +f^{fea}{}_bf^{cbg}{}_d+f^{*gaf}{}_bf^{ceb}{}_d+f^{*age}{}_b f^{cfb}{}_d=0\ .
    \end{equation}
    Here we assume a general form of a
    3-algebra, for which  structure constants are a priori
      antisymmetric only in the first 2 upper
    indices \ci{bag,che}.

    It was shown in \cite{gu,bg2} that when $f^{abc}{}_d$ are real and antisymmetric
    in $a,b,c$,  one  can construct a set of 3d
    equations of motion that are invariant under 16 supersymmetries and $SO(8)$
    R-symmetry. Adding an assumption of existence of  an
    invariant
    inner product
    \be h^{ab} = \langle T^a,T^b\rangle \ ,  \la{ina} \ee
      allows one to  construct  the Chern-Simons-matter
    $\N=8$ superconformal  BLG action
    % Lagrangian with 16 supersymmetries
    % and $SO(8)$ R-symmetry
    \cite{bg2}. This applies, in particular, for the
    $SU(N)$   Lorentzian BLG theory of \ci{gom,ben,hom}.

    This construction was further generalized in \cite{bag} to the case when the
    structure constants
     which are no longer real and antisymmetric in all
    three indices but are  required to satisfy (in addition to \rf{FI}) the following condition
    \begin{equation}\label{Riemann}
        f^{abcd}=-f^{bacd} =-f^{abdc}= {f^*}^{cdab}\ ,\ \ \ \ \ \ \ \ \ \  f^{abcd}\equiv f^{abc}{}_e
	h^{de} \ .
    \end{equation}
    It was  shown in \cite{bag} that such   more general algebras lead, in particular,
     to the ${\cal N} = 6$ Chern-Simons-matter  theory  of  ABJM \cite{aha}.

\

    Let us now review the Lagrangians of the two
    theories we are interested in.
    The
    ABJM theory  \cite{aha,benak}
   is invariant under  24 supercharge generators  (${\cal N}=6$ superconformal symmetry), $SU(4)$ R-symmetry,
      and a $U(1)$ internal symmetry.
      The field content is given by the $U(N)_{k}\times U(N)_{-k}$  CS
      gauge field $(A_\mu^{(L)}, A_\mu^{(R)})$
      and  bi-fundamental ($N\times N$ matrix-valued)  matter fields --
     4  complex  scalar fields
      $Y^A $ ($A=1,2,3,4$),
       and their  hermitian conjugates $Y^{\dagger}_{A}$, as well as
       the fermions $\psi_{A}$ and their hermitian  conjugates $\psi^{A\dagger} $.
        Fields with raised $A$
      index transform in the $\bf{4}$ of the R-symmetry
       $SU(4)$ group  and those with lowered index transform in
       the $\bar{\bf{4}}$.
      % The matter fields take values in the 3-algebra
	The corresponding Lagrangian  has the following form
    \begin{eqnarray}
        \nonumber
        {\cal L}_{_{\rm ABJM}} &=&{\cal
            L}_{_{\rm CS}} (A)   -\tr(D_\mu Y^{A} D^\mu Y^{\dagger}_A)  -V(Y)-
	    i\tr(\bar\psi^{A\dagger}\gamma^\mu D_\mu\psi_A) \no  \\	
	& & \  -\ i\frac{2\pi}{k}\tr(\bar \psi^{A\dagger} \psi_{A}Y^{\dagger}_B Y^B-
	    \bar\psi^{A\dagger} Y^B Y^{\dagger}_B\psi_{A})
        +2i\frac{2\pi}{k} \tr(\bar\psi^{A\dagger}\psi_{B}Y^{\dagger}_A Y^B-\bar\psi^{A\dagger}
            Y^B Y^{\dagger}_A\psi_{B})  \no \\
	 & & \
	    +\ i\frac{2\pi}{k}\varepsilon_{ABCD}\tr(\bar\psi^{A\dagger} Y^C\psi^{B\dagger} Y^D  )
            -i\frac{2\pi}{k}\varepsilon^{ABCD}\tr(Y^{\dagger}_D\bar \psi_A Y^{\dagger}_C\psi_B)\ . \la{ab}
    \end{eqnarray}
    Here ${\cal L}_{_{\rm CS}}$ is a Chern-Simons term and
      $V(Y)$ is a sextic scalar potential
    \begin{eqnarray}
    {\cal L}_{_{\rm CS}} &=& \frac{k}{4\pi} \epsilon^{\mu\nu\lambda}\tr\Big[A^{(L)}_{\mu}\partial_{\nu}
	A^{(L)}_{\lambda} +
            \frac{2i}{3}A^{(L)}_{\mu}A^{(L)}_{\nu}A^{(L)}_{\lambda} - A^{(R)}_{\mu}\partial_{\nu}
	    A^{(R)}_{\lambda} -
            \frac{2i}{3}A^{(R)}_{\mu}A^{(R)}_{\nu}A^{(R)}_{\lambda}\Big]\ , \la{cs}\\
     V(Y) &=& -\frac{4\pi^2}{3k^2}\tr\Big(Y^AY^{\dagger}_AY^B Y^{\dagger}_BY^C Y^{\dagger}_C + Y^{\dagger}_AY^A Y^{\dagger}_BY^B Y^{\dagger}_CY^C \no \\
    & &\ \ \ \  \ \ \ \   \ \ \ \ +\
	 4Y^A Y^{\dagger}_BY^C Y^{\dagger}_AY^B Y^{\dagger}_C - 6Y^A Y^{\dagger}_BY^B Y^{\dagger}_AY^C Y^{\dagger}_C\Big)
        \ . \la{vv}
    \end{eqnarray}
    This action is expected  to
    provide a low-energy description of  $N$ M2-branes at an  $\C^4/\Z_k$ orbifold  singularity \ci{aha}.

    In \cite{bag} the general form for the action  of a
    3d scale-invariant field theory
    with ${\cal N}=6$ supersymmetry, $SU(4)$
    R-symmetry and $U(1)$ global symmetry was found by  starting with a  3-algebra in
    which the triple product is not antisymmetric.
    The field
    content is the same as described
    above (with  the matter fields taking   values in the 3-algebra, i.e.
    $Y^A= T^a Y^A_a, \ \psi_A= T^a \psi_{Aa} $, while the gauge field is
    $A_{\mu} = A_{\mu ab} T^a \otimes T^b$) and the corresponding Lagrangian  may be  written as
    \begin{eqnarray}
        \nonumber
        {\cal L} &=&{\cal L}_{_{\rm CS}}
	 -\langle D^\mu Y_A, D_\mu Y^A\rangle-V
	  - i\langle\bar\psi^A,\gamma^\mu
	D_\mu\psi_A\rangle \no  \\
	& &\ \ \ \ \
	 -\ i\langle\bar\psi^A, [\psi_{A},Y^B;Y_{B}]\rangle
        +2i\langle\bar\psi^A,[\psi_{B},Y^B;Y_{A}]\rangle\no \\
	& & \ \ \ \  +\ \frac{i}{2}\varepsilon_{ABCD}
	\langle\bar\psi^A,[ Y^C,Y^D;\psi^B]\rangle-\frac{i}{2}\varepsilon^{ABCD}\langle Y_D,
	[\bar\psi_{A},\psi_B;Y_{C}]\rangle\ ,\label{nic}
    \end{eqnarray}
    where the brackets stand for the inner product in 3-algebra \rf{ina},
    $D_\mu Y^A_a = \del Y^A_a -    A_{\mu bd}  f^{cdb}{}_a Y^A_c$
     and
    \begin{eqnarray}{\cal L}_{_{\rm CS}}	
    &=&  \frac{1}{2} \epsilon^{\mu \nu \lambda}
\Big( f^{abcd} A_{\mu cb} \del_\nu A_{\l da} + { 2 \ov 3}  f^{acd}{}_g f^{gefb} A_{\mu ab} A_{\nu dc }
A_{\l fe} \Big) \ ,  \\
        V &=& \frac{2}{3}\langle\Upsilon^{CD}_B,\Upsilon^B_{CD}\rangle\ , \ \
        \Upsilon^{CD}_B = [Y^C,Y^D;Y_B]-\frac{1}{2}\delta^C_B[Y^E,Y^D;Y_E]+\frac{1}{2}
	\delta^D_B[Y^E,Y^C;Y_E] \no
    \end{eqnarray}
    Choosing a particular matrix realization of the  3-algebra
     such that for any 3 elements
    %for (both bosonic and fermionic) fields
     $X_1,X_2,X_3$  one has \ci{bag}
     %where $X = X^AT^A$ with $T^A$ being the generators of the 3-algebra
    \begin{equation}\label{thr}
        [X_1,X_2;X_3] = \k (X_1 X_3^{\dagger}X_2-X_2X_3^{\dagger}X_1)\ ,
	\ \ \ \ \ \ \    \ \k\equiv \frac{2\pi}{k} \ ,
    \end{equation}
    and an inner product given by an ordinary matrix trace
    \begin{equation}\la{proa}
        \langle X_1,X_2\rangle = \tr(X_1^{\dagger}X_2)\ ,
    \end{equation}
    one observes that  \rf{nic} becomes the same as the ABJM Lagrangian (\ref{ab}).
      The advantage of this form of the action \rf{nic} is that its structure
    is completely  determined by the underlying  3-algebra.

    \

    In the particular case of the totally antisymmetric 3-algebras
    the corresponding  action
    %was constructed
       \cite{bg2} has the enhanced ${\cal N}=8$ supersymmetry.
       While there is only one non-trivial
      antisymmetric 3-algebra with  a Euclidean metric \ci{papa},
       choosing a Lorentzian
      metric one finds an infinite class of 3-algebras with an underlying
       Lie algebra structure  \ci{med}.
       For any Lie algebra ${\cal G}$
    \begin{equation}
        [T^i,T^j] = f^{ij}{}_kT^k
    \end{equation}
    with structure constants $f^{ij}{}_k$ and Killing form $h^{ij}$ one can define the  corresponding
     3-algebra as follows. Let the generators $T^a$  of the
    3-algebra be denoted by ${T^-,T^+,T^i}$ $(a= +,-,i; \ i=1,\ldots,\dim{\cal G}$), where $T^i$ are in
    one-to-one correspondence with the generators of
    the Lie algebra. Then the basic 3-algebra relations are chosen to be
    \begin{eqnarray}
       [T^-,T^a;T^b] = 0\ , \ \ \ \ \ \ \
        [T^+,T^i;T^j] = f^{ij}{}_kT^k\ , \ \ \ \ \ \
        [T^i,T^j;T^k] = -f^{ijk} T^-\ ,  \la{ltr}
    \end{eqnarray}
    where $a,b$ take $(+,-,i)$  values and $f^{ijk}\equiv f^{ij}{}_lh^{lk}$ is totally
    antisymmetric, i.e.  this 3-algebra is
    antisymmetric.
    The invariant  inner product is defined as follows
    \begin{eqnarray}
        \nonumber
        &&\langle T^-, T^- \rangle = 0\ ,\ \ \ \ \
	\ \langle T^-, T^+ \rangle = 1\ , \ \ \ \  \ \langle T^-, T^i \rangle = 0\ , \\
        &&\langle T^+, T^+ \rangle = b\ ,\ \ \ \ \ \ \langle T^+, T^i \rangle = 0\ ,\ \ \ \ \ \
	\langle T^i, T^j \rangle = h^{ij}\ .
   \la{pro}
    \end{eqnarray}
    Here  $b$ is an arbitrary constant;
    since redefining  $T^+\rightarrow T^+ + a  T^-$ preserves the 3-algebra
    structure but shifts $b= \langle T^+, T^+ \rangle \rightarrow b - 2a  , $
    %\end{equation}
     we can always choose $b=0$.

    Using  this Lorentzian  3-algebra in the  ${\cal N}=8$ supersymmetric
    BLG construction (now with real scalar  fields $X^I$\  ($I=1,...,8$),
    fermions  and gauge fields  carrying  3-algebra  indices $+,-$ or $i$, where
     we may choose  $i=1, .., N^2-1$ to be
      the index  the adjoint representation of the $SU(N)$ Lie algebra)
    one finds  the following Lagrangian \ci{gom,ben,hom}\foot{Here we set
      $A^i_\mu =  A_{\mu i +}$, $B^i_\mu = \ha f^{ijk} A_{\mu jk}$  and used that
     $A_{\mu +-}$ and $A_{\mu -i}$ decouple.
    % ($h_{ij} = \delta_{ij}$).
     We  then
      replaced the fields with adjoint index $i$  with matrices
      in the fundamental representation of $SU(N)$
      (i.e.   $X^I_i$, etc,  by $N \times N$ hermitian matrices)
      and thus replaced  summation over $i$ (with $h_{ij} = \delta_{ij}$)
      by the trace.}
    \begin{eqnarray}
        \nonumber
        {\cal L}_{_{\rm L-BLG}} &=&
	\tr\Big[
	-\frac{1}{2}\big[D_{\mu}(A) X^I-B_{\mu}X^I_+\big]^2+
	\frac{1}{4}(X_+^K)^2 ([X^I,X^J])^2 - \frac{1}{2} (X_+^I [X^I,X^J])^2 \\
	 & & +\ \frac{i}{2} \bar{\Psi} \Gamma^{\mu} D_{\mu}(A)  \Psi + i \bar{\Psi}_+
	\Gamma^{\mu} B_{\mu} \Psi -\frac{1}{2}\bar{\Psi}_+ X^I
        [X^J,\Gamma_{IJ}\Psi] + \frac{1}{2}\bar{\Psi}X^I_+[X^J,\Gamma_{IJ}\Psi]\no  \\
	\la{lor}
        & & +\ \frac{1}{2} \epsilon^{\mu\nu\lambda} F_{\mu\nu} B_{\lambda} -
	 \partial^{\mu} X^I_+\  B_{ \mu} X^I \Big]
	 %+ {\cal L}_{gh}\ ,\la{lor}
    % \end{eqnarray}
    % where $A^i_\mu =  A_{\mu i +}$, $B^i_\mu = \ha f^{ijk} A_{\mu jk}$
    % while $A_{\mu +-}$ and $A_{\mu -i}$ decouple ($h_{ij} = \delta_{ij}$),  and
    % \begin{equation}
     %    {\cal L}_{gh} =
     + \tr\Big[ -
	%\tr\Big[
	 \partial^{\mu}X^I_+ \partial_{\mu}X^I_{-} + \frac{i}{2}
	\bar{\Psi}_{-}\Gamma^{\mu}\partial_{\mu}\Psi_+
	\Big]
	\ .
	% \la{gho} \end{equation}
     \end{eqnarray}
    This theory having the right symmetries and an  arbitrary-rank $SU(N)$  gauge group
    was proposed \ci{gom,ben}  as a candidate for a low-energy limit of a
    theory on $N$ M2-branes in flat space.\foot{In contrast to similar plane-wave sigma models
    where the existence of a negative  norm direction in field space is natural and
    does not represent a problem due to reparametrization invariance allowing to fix a light-cone gauge,
    here there is an  apparent non-unitarity issue.  It is possible  that  it can be
    avoided  by a consistent truncation of the spectrum or by  gauging
    the shifts in  $X_-$ direction \ci{sch,gomi}.}

    Like  similar models in two  \ci{nap} and higher \ci{tse2} dimensions
     it has  a scale symmetry
    that allows one  to absorb the coefficient in front of the Lagrangian  into a field redefinition.
     It thus
    has no obvious coupling constant and thus no natural  perturbative expansion
    preserving all the symmetries
    (if one chooses a particular expansion point in field space one can use the
    standard loop expansion but
    a non-zero  background   would spontaneously break the conformal invariance).

    Assuming one is allowed to consider only the observables that do not depend on $X_-$,
    one may integrate over $X_-$  getting a delta-function constraint $\del^2 X_+=0$.
     Solving it by $X_+ = {v}$=const
      and integrating out $B_{\mu}$ one  obtains   \ci{gom}--\ci{mukh2},
      via simple path integral duality  transformation \ci{mukh1},  the 3d  $SU(N)$
   $\N=8$ SYM  theory with the coupling constant $g_{YM} = {v}$, i.e.
   one establishes a relation between the L-BLG  theory and the low-energy theory
    of $N$ coincident  D2-branes.\foot{In a sense,   the L-BLG theory can be interpreted as
    a  ``conformal  dressing'' of the 3d $\N=8$ SYM theory where one effectively
       integrates over its  coupling constant.
      This does not,  however, mean that the two theories are   completely equivalent,
       cf. footnote 1  and a discussion below. }

     In principle, this does not necessarily  need to imply   that
    the L-BLG theory \rf{lor} is equivalent to D2-brane theory --
    this is so only if it is treated  in perturbative expansion  near the vacuum $X_+ = {v}$=const
    which spontaneously breaks conformal symmetry and introduces $v$ as a coupling constant.
    To preserve the conformal symmetry  one may insist on summing over all  solutions of
    $\del^2 X_+=0$, and,  in particular, on integrating over $v$  (cf. \ci{gomi,ver,hon}).
    It remains to be seen if one  can give a  precise meaning to  such
    ``summation'' \ci{ver} or ``integration over
    coupling constant'' (which obviously depends on a definition of a measure of integration, etc.).

   \

   Our  aim below  will be to make precise, following   the  suggestion of \ci{hon},
    in which sense  the L-BLG theory
   \rf{lor} can be interpreted as a  scaling limit of the ABJM theory \rf{ab}.
   One may  hope  that understanding the L-BLG theory via this scaling limit
   may possibly shed light on its  proper interpretation
    and  suggest an alternative  way of defining the  quantum L-BLG   theory
  without explicit breaking of its conformal invariance.
  As we shall see,
  to  implement such  a scaling limit in a systematic fashion one will need
  to supplement the ABJM theory by an extra ``ghost'' supermultiplet,
  and thus it may not help with clarifying the unitarity issue of L-BLG theory.

  %%%%%%%%%%%%%%%%%%%%%%%%%%%%%%%%%%%%%%%%%%%%%%%%%%%%%%%%%%%%%%%%%%%%%%%%%%%%%%%%%%%%%%

\section{Scaling limit}
% of gauge algebra and of  the action}
% 3-algebra}

Here  we shall first review  the scaling limit of Chern-Simons theory and
then show how
a similar limit can be used to relate the ABJM action to  L-BLG action
  by first adding an extra singlet ghost  multiplet to the  ABJM   theory.

Next, we shall show
 that there exists a  scaling limit or contraction  of the
  non-antisymmetric 3-algebra (\ref{thr}) associated to the $
  U(N) \times U(N)$ gauge group
   \ci{bag}  supplemented by an extra  negative-norm  generator  that
     reduces it to the
      antisymmetric  Lorentzian 3-algebra associated to the $SU(N)$ gauge group.
     This will   provide  a  rigorous  confirmation  of the relation  between the full  non-linear
      ABJM and  L-BLG actions via the scaling limit.

\subsection{Examples of scaling limits of  Chern-Simons theory}

Let us  start with recalling a simple example of a relevant type of
scaling limit -- the one that leads from the   $U(1) \times SU(2)$ algebra
to  the centrally extended Euclidean algebra in 2 dimensions $E_2^c$.
The  idea \ci{sfets}  will be to mix together the $U(1)$ generator $J^0$  with a
Cartan $U(1)$ generator $J^3$ of  $SU(2)$.
Denoting  the remaining two generators of $SU(2)$   by   $J^n$,\ $n=1,2$
we have
\be
 [J^0,J^n]=0,\ \ \  [J^0,J^3]=0, \ \  \ [J^n,J^m] =  \ep^{nm} J^3 ,\ \ \ [J^3, J^n] =
    \ep^{nm} J^m  \ .\la{kj} \ee
Let us introduce $\JJ^+,\JJ^-,\JJ^n$ by setting
\be \la{ukj}
J^0= \g^{-2} \JJ^-,\ \ \ \ \ \ J^3 = \JJ^+ + \g^{-2} \JJ^-,\ \ \ \ \ \ J^n = \g^{-1} \JJ^n\ , \ee
  and take the limit  $\g\to 0$.
 Then we  end up with the algebra   of  $E_2^c$:\foot{ An equivalent (up to $J_0 \to - J_0$ and shift of generators)
   prescription leading to the same algebra  is to define
  $
J^0=  \ha \JJ^+  -  \g^{-2} \JJ^-    , $ $
 \  J^3 = \ha \JJ^+ + \g^{-2} \JJ^- ,$ $   \ J^n = \g^{-1} \JJ^n\ , $
  i.e. $ \JJ^+=  J^3 + J^0, \ \JJ^- =  \ha \g^2 ( J^3 -J^0)$.
  Then $[\JJ^-, \JJ^n]=0$  and we get back  the rest of \rf{uj}.}
\be  [\JJ^-,\JJ^n]=0,\ \ \  \ \ \ [\JJ^-,\JJ^+]=0, \ \  \ \  \ \
[\JJ^+, \JJ^n] =   \ep^{nm} \JJ^m\ , \ \ \ \ \
  \ [\JJ^n,\JJ^m] =  \ep^{nm} \JJ^- \ .\la{uj} \ee
  If we consider the corresponding   quadratic form (or Casimir)
  on the original algebra  taking the $U(1)$ part with negative sign
  we get
  $- J^0 J^0 + J^3 J^3 + J^n J^n =\g^{-2} ( 2   \JJ^-  \JJ^+   + \JJ^n\JJ^n)$
   which is  non-degenerate  after  an overall rescaling.
While the standard Killing form on $E^c_2$ is degenerate,
it admits a  non-degenerate invariant bilinear form with the
signature  $(-1,1,1,1)$  with   entries
\ci{nap,can}
\be \langle \JJ^m,\JJ^n\rangle =  \delta_{mn}, \ \ \ \
\ \langle \JJ^m,\JJ^\pm\rangle = 0 , \ \ \
 \ \  \langle \JJ^-,\JJ^-\rangle =0, \ \ \
\langle \JJ^-,\JJ^+\rangle =1,\ \ \  \langle \JJ^+,\JJ^+\rangle =b, \la{guj}  \ee
where $b$ can be set to 0 by a redefinition of generators.
This  may be compared to \rf{ltr},\rf{pro} where $T^-$ corresponds
to the central element $\JJ^-$ and
$T^+$ -- to the  redefined  rotation generator $\JJ^+$.

As discussed  in \ci{tse2}, the CS theory  for $E^c_2$ with the above
non-degenerate invariant form
can be obtained by a scaling  limit from CS theory for $U(1)_{-k} \times SU(2)_k  $:
  one should do  a   redefinition  of  gauge field  components  and take
$k \to \infty$.
 Explicitly,
starting with
\be
S_{_{U(1)_{-k} \times SU(2)_k}}
=  { ik\ov 8\pi}  \int  d^3 x \  \ep^{\m\n\l}
\big(  A_\m{}_0 \del_\n  A_\l{}_0  -   A_\m{}_3 \del_\n  A_\l{}_3   -
 A_\m{}_n \del_\n  A_\l{}_n
-    \ep_{nm}  A_\m{}_n A_\n{}_m A_\l{}_3  \big) ,  \la{pp}
\ee
and setting
\be \la{prp}
A_\m{}_0 =  -\AA_\mu{}_+ + \g^2   \AA_\mu{}_-   \ , \ \ \ \
A_\m{}_3 =  \AA_\mu{}_+    \ ,\ \ \ \
A_\m{}_n = \g \AA_\mu{}_n  \ , \ \ \ \ \ k= \g^{-2} \td k \ , \ee
and then taking the limit $\g \to 0$  (with $\AA_a$ and $\td k$ fixed)
we end up with
\be
S_{_{E^c_2}} = - { i\td k\ov 8\pi}  \int  d^3 x \  \ep^{\m\n\l}
\big( 2  \AA_\m{}_- \del_\n  \AA_\l{}_+    +  A_\m{}_n \del_\n  A_\l{}_n
+    \ep_{nm}  A_\m{}_n A_\n{}_m \AA_\l{}_+   \big)  \ , \la{gj}
\ee
which is the CS action for $E^c_2$  with the metric \rf{guj} with $b=0$.\foot{
In general, the  CS  Lagrangian  written in terms of an invariant metric
$\Omega_{ab}$ and structure constants of the  algebra $f^{a}{}_{bc}$ is
$L = \Omega_{ad} \ep^{\m\n\l}
\big( A^a_\m \del_\n  A^d_\l  + { 1 \ov 3}
 f^{a}{}_{bc}  A^d_\m A^b_\n  A^c_\l\big)$.}

 The redefinition in  \rf{prp} is consistent with the one \rf{ukj}
  used in taking the scaling limit in the algebra, namely,
 \be \la{redq}
 A_0 J^0 + A_3 J^3 + A_n J^n = \AA_+ \JJ^+ + \AA_- \JJ^-  + \AA_n \JJ^n  \ . \ee
Note that the  action \rf{gj} does not actually depend on $\td k$ as it can be redefined
 away by rescaling the fields. This feature is shared by the L-BLG    action  --
 the overall coefficient in \rf{lor} can be set  equal to 1  by a field redefinition
  \ci{gomi}.\foot{One is to rescale
 the fields in \rf{lor} as follows:
 $X^I_i \to \k  X^I_i, \ X^I_+ \to \k^{-1}  X^I_+,\ X^I_- \to \k^3  X^I_-,\ B \to \k^2 B$, \
\ $\k= k^{-1/2}$.}

 \

 Let us  mention also  another
  example of a limit  of CS   theory that leads to a BF theory   (cf. \ci{witt}).
 In this case we   start with CS theory  for the group $G_{-k} \times G_k$, e.g.,   $G= U(N)$.
 %in the ABJM case.
 Let us  denote the two  gauge fields  as $A^{(L,R)}_\m$
 and define their combinations $A_\m$ and $B_\m$  as
 \be
 A^{(L)}_\m = A_\m -  \ha \g  B_\m  \ , \ \ \ \ \  A^{(R)}_\m = A_\m + \ha  \g  B_\m  \ ,
 \ \ \ \ \ \ k =  \g^{-1} \td  k \ .\la{defq}
  \ee
  Then taking the limit $\g \to 0$ (i.e. $k\to \infty$ for fixed $\td k$)
   in the corresponding action   we end up with (cf. \rf{gj})
  \be
S = - { i\td k\ov 8\pi}  \int  d^3 x \  \ep^{\m\n\l}    B_\m F_{\n\l} (A)     \ . \la{gjk}
\ee
 Here $\td k$ can be set to 1 by a redefinition of $B_\m$.

  The corresponding limit at the level of gauge algebra can be defined  as follows:
  if the generators of $G \times G$ are   $ T^{(L)}_i, T^{(R)}_i$  with
  both sets satisfying $[T_i, T_j]= f_{ij}{}^k T_k$,   then we may define
  \be
  T_i =  T^{(R)}_i + T^{(L)}_i\ , \ \ \ \ \ \
  P_i =  \ha \g ( T^{(R)}_i  -  T^{(L)}_i)
  % T^{(L)}_i = \ha T_i +   \g^{-1}  P_i  \ , \ \ \ \ \  T^{(R)}_i =  \ha T_i - \g^{-1}   P_i
    \ .
  \ee
  In the limit $\g \to 0$
  we will then  get  a  semidirect sum of the algebra of $G$ and translations, i.e.
  \be
  [T_i, T_j]= f_{ij}{}^k T_k\ , \ \ \ \ \ \  [P_i, T_j] = f_{ij}{}^k P_k\ ,  \ \ \ \ \
   [P_i, P_j] = 0   \ .  \ee
  This relation of  the $BF$ structure  of the CS part of the L-BLG action \rf{lor}
  to such Inonu-Wigner contraction-type scaling limit
   was  mentioned earlier  in \ci{gom,ben,hon}.

 \subsection{From   ABJM + ``ghost'' action to  L-BLG action\\  via a scaling limit}

We are now ready  to  discuss   the scaling limit of the ABJM action
 suggested in \ci{hon}.
We shall concentrate on the bosonic part of the action
(the fermions are readily included after  we  describe  the  corresponding scaling limit at the level
of the 3-algebras  below).  In addition to the
$U(N)_{-k} \times U(N)_k$   CS  action the
bosonic part of the ABJM action \rf{ab} contains also the ``matter'' part --
complex $N \by N$  scalar  field matrix
$Y^A$  ($A=1,2,3,4$)  in bi-fundamental   representation of $U(N) \times U(N)$
with the kinetic term $ - \tr ( D_\mu Y^A   D^\mu   \bY_A )$, where   $\bY_A= (Y^A)^\dagger$.
Here
\be \la{cova}
D_\m  Y^A = \del_\m  Y^A  +  A^{(L)}_\m Y^A  -  Y^A A^{(R)}_\m
= \del_\m  Y^A  +  [A_\m,  Y^A ]   - \ha  \g  \{B_\m, Y^A\}  \ , \ee
where   we used the same field redefinition as in \rf{defq}.
If we  now  separate the trace part in  $Y^A$ by setting
\be\la{pt}
 Y^A = Y^A_0  \oI   + \td Y^A , \ \ \ \ \
  Y^A_0= \g^{-1} Y^A_+ \ , \ \ \ \ \   \tr\  \td Y^A =0   \ ,  \ee
 we get  4 complex singlet fields $ Y^A_+  $ and $4(N^2-1)$   complex  scalar components of
 $\td Y^A$. We have introduce  the factor of $\g^{-1}$ in the singlet part to define
 the scaling limit leading  to the L-BLG action.  Then we get
 \be \la{coo}
&& -\tr |D_\mu Y^A |^2
 =- \tr   |\td D_\mu  \td Y^A  |^2
 %\td  D^\mu    \td Y_A )
 +   \tr   | \g^{-1}\del_\m  Y^A_+  - \ha  \g  \{B_\m, \td Y^A\}|^2          \ , \\
&& \ \ \
  \td D_\mu  \td Y^A \equiv  \del_\m  \td Y^A   +  [A_\m,  \td  Y^A ]  -
       B_\m Y^A_+    \ . \ee
 Taking the limit  $ \g \to 0$  the second term  in \rf{coo} gives
 \be
 - \tr | \g^{-1} \del_\m Y^A_+  - \ha  \g  \{B_\m, \td Y^A\}|^2 \to
 -  N \g^{-2} |\del_\m Y^A_+|^2    +  2  \big[\del_\m Y^A_+ \tr  (B_\m \td Y^A)  + c.c. \big]  \ .
 \la{sii} \ee
 As was  observed in \ci{hon},
  the  first term  in \rf{coo} and the second term in \rf{sii}  are
   as in the L-BLG   action \rf{lor}   with the 4 complex  singlet  scalars
    $Y^A_+$ corresponding to   the  8 real scalar fields $X^I_+$; the same was found
     to be
     true also for the bosonic  interaction terms \ci{hon}.\foot{The $U(1)$
    component of the $U(N)$ field
    $A_\m$  field automatically decouples, while  the $U(1)$  component of the $U(N)$ field
    $B_\m$  field  can be decoupled by rescaling it by an extra factor of $\g$.}

    \

 One  is still left   with a singular first term in \rf{sii}, $\sim
 \g^{-2} |\del_\m Y^A_+|^2$.   In  \ci{hon} it was suggested
 that to make the action finite
 one is to  require  that $\del^2 Y^A_+=0$  which was then concluded to be the same
 as the equation  in L-BLG theory \rf{lor} obtained by  variation over  $X^I_-$.

 This  suggestion, however, appears to be hardly satisfactory for several reasons:

 (i) The condition for the vanishing of that singular term is,  in general,  ambiguous
 (depending, e.g., on boundary conditions)  and, in fact,   appears to require
 that $\del_\m Y^A_+=0$. Then $Y^A_+$=const but   in this case the L-BLG action  becomes
  equivalent to 3d SYM or  D2-brane action. The  relation  to the
  ABJM theory via a scaling limit
  with $k \to \infty$ is then not too  surprising as the  D2-brane theory is
  also  a limit of the ABJM theory
   \ci{aha,pang}. The limit considered in \ci{hon} then
   does not represent a consistent  derivation of the L-BLG theory from the
   ABJM one but rather of a
   D2-brane theory from  the ABJM theory.

 (ii) The non-trivial difference of the \lb theory  from  D2-brane theory
 is in the presence of  extra 8 scalar fields $X^I_-$   on which ``observables''
 (composite conformal operators
 representing states) may, in general,  depend  and which may
  enter the external sources (cf. \ci{ver}).
 However,  there is no place for such fields in the above  limit of the  ABJM theory
 considered in \ci{hon}:
 the matter  fields $Y^A$ or  $ Y^A_+, \td Y^A$   correspond  only to $X^I_+, X^I$.

 (iii) The above  limit  missing $X^I_-$ fields  is also not consistent with
 an  expectation that the scaling limit may be
 may be carried out directly  at the level of the corresponding 3-algebras:
 the Lorentzian 3-algebra  contains the  generator $T^-$ corresponding to
 $X_-$    (in addition to  the generators corresponding to  the $SU(N)$ algebra
 and the $X_+$ fields)  but   the 3-algebra for the ABJM theory  lacks the
 corresponding  generator.
  A related point is that  while  the scalar  product on the
 Lorentzian 3-algebra vector space is indefinite,  the scalar product on the ABJM 3-algebra
 vector space is hermitian (reflecting the positivity of the scalar kinetic term
 in the  ABJM action).

 Here we are going to improve    the suggestion  of
   \ci{hon} by  proposing     that in order  to  obtain the L-BLG theory by
 a consistent scaling limit  similar to the one that relates the corresponding CS parts of the
 actions  one is to extend the  original ABJM theory by an extra ``ghost''
  $U(1)$  multiplet  containing
 4 complex bosonic fields $U^A$ and the corresponding singlet fermions.
 To implement such  scaling limit   at the 3-algebra level one will then
  need  to extend
 the ABJM 3-algebra by an extra  negative-norm generator.

  Namely, we shall start with  the (bosonic) ABJM Lagrangian
  supplemented by an extra term $  N |\del_\mu U^A|^2$ having    ``wrong'' sign
 of its   kinetic term. Then  setting
 \be   \la{fgd}
  U^A = -\g^{-1}  Y^A_+   +    \g  N^{-1}  Y^A_-  \ , \ee
  where  $ Y^A_+ $ is defined in \rf{pt} and
   $ Y^A_- $ is a new variable, and taking the limit  $\g\to 0$
   we shall then  get instead of \rf{sii}
  \be
    N |\del_\mu U^A|^2  -  \tr | \g^{-1} \del_\m Y^A_+  -  \ha  \g  \{B_\m, \td Y^A\}|^2
     \  \to \
    - \del_\mu  Y^A_+ \del^\mu  Y^{A}-   +  2 \big[  \del^\m Y^A_+ \tr  (B_\m \td Y^A)  + c.c.  \big]
 \la{pii} \ee
 As a result, we  recover the full kinetic term of the L-BLG theory  \rf{lor}
 (with $X^I_\pm$ being  the real parts of $Y^A_\pm$).

  This  limiting procedure
  is obviously similar to the one discussed  above on the example of
 the $U(1)_{-k} \times SU(2)_k$  CS model
 or the one used
 in the  string sigma model  context in getting a pp-wave model via a Penrose-type limit
 \ci{sfets} (with $U^A$ playing the role of the ``target space time'' direction).

 With a hindsight, the  need to extend the ABJM action by an  extra  ``ghost'' $U^A$ field
  is  hardly unexpected: it would be strange to obtain
 the  L-BLG
 action which  has an indefinite kinetic-term signature
   from a manifestly definite  ABJM action  by a regular   scaling limit.
The existence  of this scaling limit   does  not seem to shed extra light on the unitarity issue
 of the L-BLG    theory:\foot{Also, it does not seem possible to obtain the gauged
 version \ci{sch,gomi} of the \lb theory  by a scaling limit of  (a version of) the ABJM theory.}
  while at the level of the ABJM theory  one may assume that the  extra ghost multiplet is
 completely  decoupled (and does not enter the observables),
 it gets   effectively coupled   via the redefinition   \rf{fgd}
 in the process of  taking the scaling limit.

 \

 \subsection{From ABJM + ``ghost''   3-algebra to L-BLG 3-algebra\\ via a scaling limit}
% Derivation of L-BLG action from  ABJM one}

 Let us now  show
 that starting   with the 3-algebra (\ref{thr}) for  the  ABJM theory extended by
  an extra ``ghost'' generator  one can get the
     Lorentzian 3-algebra \rf{ltr} associated to L-BLG theory  by a  scaling limit.
     Since the two supersymmetric actions  can be  directly constructed from the
     corresponding 3-algebras  \ci{gom,ben,bag},
     that will imply, in particular,  that the scaling limit
     will go through  at the level of  the full actions including  the fermions  and
     all interaction terms.

% Here we shall use logic of \ci{mede}.
 The Lorentzian 3-algebra vector space has an indefinite scalar product \rf{pro}  while
  that of the ABJM algebra \rf{proa}
   is positive definite. Counting  the degrees
   of freedom in the  $U(N)\times U(N)$ ABJM theory we get  $4\times2N^2 = 8N^2$ scalars,
    while in the $SU(N)$ L-BLG  theory we get  $8\times(1+1+(N^2-1)) = 8N^2 + 8$ scalars.
     To match the degrees of freedom we
      thus  need to add an extra generator to the ABJM 3-algebra;
      it should have   a negative
     norm and should thus  correspond to the ``ghost" multiplet  introduced
     in  the previous subsection.

Let us  start by considering the scaling limit for the 3-algebra corresponding to the
 simplest
case of the $U(2)\times U(2)$ gauge group. We decompose the $2\times2$ complex matrices $X$
in \rf{thr} (which may be identified with maps  from $\C^2 $ to $\C^{*2}$, or elements of
 bi-fundamental representation of   $U(2)\times U(2)$ \ci{bag}) in  the following basis
 (with complex coefficients): $ \{E,\sigma^{i}\}$. Here  $E=iI_2$ with $I_2$ as the
  $2\times 2$ identity matrix and $\sigma^{i}$ are the Pauli matrices. The 3-algebra relations
   (\ref{thr}) for these  basic elements then become
    \begin{eqnarray}\label{TA}
        \nonumber
        &&[E,\sigma^i;\sigma^j]=[\sigma^i,\sigma^j;E]=2\k \epsilon^{ijk}\sigma^k \ , \\
        &&[\sigma^i,\sigma^j;\sigma^k]=  -2\k  \epsilon^{ijk} E
	\ , \ \ \ \ \ \ \ \ \     \k=  \frac{2\pi}{k} \ .
    \end{eqnarray}
    Note that this $U(2)\times U(2)$ (or $N=2$) case is special in that the 3-algebra
     is actually fully antisymmetric. This  implies
      that in this case the corresponding action
        will have the extended $\N=8$ supersymmetry  \ci{aha}.
	%AT
	\foot{
	The reality condition implies omitting the $U(1)$ factors, 
	and then the corresponding  $SU(2) \times SU(2)$ theory is equivalent to the  BLG
	theory. In general, the issue  of $U(1)$ factors   appears to be subtle one, and 
	$U(1)$'s   may be required for a consistent description of two M2-branes on 
	the orbifold.}

 By analogy with the discussion in the previous two subsections
 % (and earlier  construction in pure CS theory)
  let us now extend the 3-algebra by
 adding  an extra generator $e$  in a trivial  way, i.e.
  without modifying the existing relations and with
   $[e, T^a; T^b] = 0$.
   We shall also assume that the scalar product on the extended algebra
   is defined so that
      $e$ is  perpendicular to the  $ \{E,\sigma^{i}\}$  generators and
      has negative norm: \
      \be \la{scap} \langle e, e \rangle = -2\ . \ee
  Similarly  to the $U(1)\times SU(2)$  CS example discussed above
    in section 3.1,  let us  now rescale $k$ and  rename the generators
    to $T^+,\ T^-,\ T^i$ as follows
    \begin{eqnarray} e=  2\g^{-1} T^{-} \ , \ \ \ \ \
     E=   \g   T^+ + 2\g^{-1} T^- \ , \ \ \ \ \ \ \sigma^i= T^i \ , \ \ \ \ \  \ \ \
     k = \g^{-1} \td k \ ,  \la{reg}
     \end{eqnarray}
     i.e. $ T^{-} = \ha \g   e, \ \ \ \ T^+ =  \g^{-1} ( E-e)$.
      Next, let us
     take the limit $\g\to 0$ for fixed $T^\pm,\ T^i,\ \td k$.
     This leads to    the following (totally antisymmetric) 3-algebra
     ($ a,b= +,-, i;$ $ \  i=1,2,3$)
    \begin{eqnarray}
        \nonumber
        &&[T^{-},T^a,T^b] = 0\ ,  \\
        \nonumber
        &&[T^+,T^i,T^j] = f^{ij}{}_kT^k\ ,\ \ \ \ \ \ \ \
	f^{ij}{}_k =  2 \td \k \epsilon^{ijk}  \ ,  \ \ \  \ \  \ \td \k\equiv  { 2 \pi \ov \td k
	} \ ,  \\
        &&[T^i,T^j,T^k] = -f^{ijk}T^{-}    \ , \ \ \ \ \ \ \ \
	f^{ijk}= 2f^{ij}{}_{n} \delta^{nk}=
	  4\td \k\epsilon^{ijk}    \ .
    \end{eqnarray}
   Here $\td \k$ can be set equal to 1  by  further  rescaling of the generators.
   This
    is just  the Lorentzian 3-algebra \rf{ltr} associated to the gauge group  $SU(2)$,
    with  \rf{scap} leading to the appropriate scalar  product \rf{pro}:
   \begin{eqnarray}
        \nonumber
        &&\langle T^-, T^- \rangle = \frac{\g^2}{4}\langle e, e \rangle  \rightarrow 0\ ,  \ \ \
	\ \ \ \ \ \ \
        \langle T^+, T^+ \rangle = \g^{-2}\langle E-e, E-e \rangle = 0\ ,  \\
        &&\langle T^+, T^- \rangle = \frac{1}{2}\langle E-e,
	e \rangle = -\frac{1}{2}\langle e, e \rangle = 1 \ . \la{scala}
   \end{eqnarray}
   In the  general case of $U(N)\times U(N)$ Lie algebra,
    the generators $T^i$ of $SU(N)$ satisfy
    \begin{equation}
        T^iT^j =- i  \delta^{ij} E  + \frac{1}{2}(if^{ij}{}_k+d^{ij}{}_k)T^k \  ,
    \end{equation}
    so that the generalization of the $N=2$
    3-algebra relations
     \rf{TA}  is
    \begin{eqnarray}\label{Tijk}
        \nonumber
        &&[E,T^i;T^j] = \k f^{ij}{}_kT^k \ ,  \\
        &&[T^i,T^j;T^k] = - N^{-1} \k f^{ijk} E + \k A^{ijk}{}_mT^m \ ,
    \end{eqnarray}
    where the coefficients $A^{ijk}{}_m$ are   determined  by  $f^{ij}{}_k$ and $d^{ij}{}_k$:
    \begin{equation}
        A^{ijk}{}_mT^m = N^{-1} h^{jk}T^i + \frac{1}{4}(if^{kj}{}_l+d^{kj}{}_l)(if^{il}
	{}_m+d^{il}{}_m)T^m - (i\leftrightarrow j)\ .
    \end{equation}
    Unlike the $N=2$ case \rf{TA} here
    there  is an  additional term in the last relation in   \rf{Tijk}
    which is not antisymmetric in $i,j,k$.
    However,  after adding an extra ``ghost'' generator $e$  that commutes with all other
     generators and  taking
     a similar
      scaling limit   with
       \begin{eqnarray}
            e = N\g^{-1}T^- \ , \ \ \ \ \ \ \ \ \
        E = \g T^+ + N\g^{-1}T^- \  , \ \ \ \ \    k= \g^{-1} \td k\ ,\ \ \ \ \ \ \g\to 0 \ , \la{ghp}
    \end{eqnarray}
     the second term in the last line of \rf{Tijk} will
      vanish and we will get the
     antisymmetric $SU(N)$ L-BLG
    algebra \rf{ltr}.
    %  like in the $U(2)\times U(2)$ case.
    Note that  the definition of the generators in \rf{ghp}
    is in direct correspondence with the definition of fields in \rf{pt},\rf{fgd}:
   % if $Y= Y_0   \oI   + \td Y $, then
   \be  Y_0 E + U e =  Y_+ T^+ + Y_-T^-   \ . \ee

\section{Concluding remarks}

 To  conclude, the 3-algebra  \ci{bag} corresponding  to the ABJM  theory is
  related to the Lorentzian  antisymmetric 3-algebra corresponding  to the  L-BLG
   theory after one   extends  the former   by a  ``ghost''   generator   and
   takes the  scaling limit defined above.  This    implies that the
   corresponding interacting actions  are also  related by
   this  scaling limit.

 One may wonder  if this  scaling-limit relation may have some  implications for the physical
   interpretation of
   the  L-BLG  theory, and the unitarity issue in particular.
   The ABJM theory extended   by an   abelian  ``ghost'' supermultiplet ($U^A$ in and
   the corresponding fermions)  may formally be unitary assuming
   we define the corresponding
   observables (conformal operators and their correlation functions)
   so that they do not depend on the ``ghost''
   fields.   The  observables of the L-BLG theory  obtained by taking
   the scaling limit  will then  be a certain subset of all  possible
   composite operators one could consider  by starting directly
   with the full L-BLG action.

   The relation via the scaling limit does not necessarily
   mean that the correlation functions of the two
   theories will also be  directly related: while the scaling limit is   smooth in the action, it need not be so
    at the level  of the correlation functions. One may
     view the  L-BLG theory as a certain ``truncation''  of the ABJM theory
   (similar to the case
     of 2d sigma models related via Penrose-type limit).

   One consistency test of this scaling limit relation
    between the ABJM   and the L-BLG
   theories  is found by considering their known connection to  3d $\N=8$
   $SU(N)$ SYM theory  describing  $N$ coincident D2 branes.
   Starting from the ABJM  theory
    one can get the theory on D2-branes by assuming that
     one of the scalars $Y^A$
      develops an
    expectation value $\langle Y\rangle = \sqrt{k} \ v $ and then taking the limit
     $k\rightarrow\infty$, $v\rightarrow\infty$ with
      $g_{\rm YM}^2={v^2\ov k}$  kept fixed \ci{aha,dis,pang}.
      Since the new ``ghost'' field $U$
      that we introduced is completely decoupled it  can be integrated out
      without changing this relation to 3d SYM theory.
       Given that the
      L-BLG also reduces to  the  D2-brane theory   when one of  the 8 scalars $X^I_+$
       gets  an expectation value, one would naturally expect that taking the
        above scaling limit and then setting $\langle X_+\rangle = \tilde{v}$ would
	be equivalent to the D2-brane reduction procedure of
	 \cite{dis,pang}.
	
	 Indeed, the scaling limit translates
	the D2-brane  limit  of assuming a scalar expectation value and
	taking it to infinity to first  scaling  the relevant fields to infinity and then
	giving one of them a finite expectation value. Comparing the
	resulting coupling constant $g_{\rm YM}$
	 for the
	 two ways of getting the $\N=8$ SYM  theory -- from the ABJM theory and from the L-BLG theory -- we see that it is the same
    \begin{equation}
        g_{\rm YM} = \tilde{v} = \langle X_+\rangle = \frac{1}{k}\langle Y \rangle =
	\frac{1}{k}\sqrt{k}v = \frac{v}{\sqrt{k}} = g_{\rm YM}\ .
    \end{equation}
    Thus  L-BLG theory may be interpreted as
    a background-independent intermediate step that can be considered
     when reducing from the
     M2-branes to the D2-branes.

The  limit  from the ABJM theory to D2-brane theory
     may be  interpreted \cite{dis} as the cone $\mathbb{C}^4/\mathbb{Z}_k$
    becoming locally a cylinder for
    large $k$, with  the radius of the cylinder being
     related to the distance of the branes from the
     singularity, given by the expectation value of the scalar
    $Y$. The scaling limit leading to the L-BLG theory
    also converts the cone into a cylinder, but in this  case
    the radius of the cylinder is a dynamical
    variable  and only when we give it an expectation value do we recover $D2$-brane
    theory.

   Let us note also that the scaling
   limit gave us  the   $SU(N)$ (and not $U(N)$) L-BLG from
   the $U(N)\times U(N)$ ABJM theory, with $Y_+$
    related to the center-of-mass mode of the branes (cf. \ci{sen}).
    This interpretation  is consistent with both the structure
     of the 3-algebras, and the shift symmetry of the action.
     The  limit scales out half of
     the fluctuations and seems to move the
     M2 branes infinitely far from the singularity, which presumably
     is why we  recover more supersymmetry
     ($\N=8$  of L-BLG instead of $\N=6$ of ABJM), but it
     also shrinks the size of the cone.
     % (apparently getting rid of one dimension).
     %, the meaning of which is unclear at the moment.

\

Given  that the ABJM theory  is dual to
 type IIA string theory on $AdS_4 \times \ICP^3$ \ci{aha}
(with parameters $g_s = N^{-1} ( N/k)^{5/4}, \  \a' = \l^{-1/2} = \sqrt{ k/N} $)
one may wonder if the scaling limit  leading to the L-BLG theory may
mean for  the dual string theory.  Naively, sending $k$ to infinity
at fixed $N$ appears to correspond to free zero-tension strings.
Since the limit involves also a particular scaling of the fields, i.e.
the  L-BLG theory corresponds to a certain truncation of the ABJM + ``ghost''  theory,
the possibility of a string-theory interpretation of this limit remains unclear.

Among  open problems let us mention also
     the   question of  existence of    other similar limits of
      non-antisymmetric
     3-algebras in the  general setting   presented in \cite{mede}.

\bigskip

\section*{Acknowledgments }
%%%%%%%%%%%%%%%%%%%%%%%%%%%
We are grateful to  J. Bedford and O. Lunin  for  useful discussions.
This work was supported in part  by the  STFC rolling grant.

\end{document}